\documentclass[aps,prb,twocolumn]{revtex4}
\usepackage{graphicx}
\usepackage{color}
\definecolor{darkred}{rgb}{0.75,0.0,0.0}
\newcommand\beq{\begin{equation}}
\newcommand\eeq{\end{equation}}
\newcommand\bear{\begin{eqnarray}}
\newcommand\eear{\end{eqnarray}}
\begin{document}

\title {Electron-Electron Interactions on the Edge States of Graphene: 
A Many Body Configuration Interaction Study} 

\author{Sudipta Dutta, S. Lakshmi and Swapan K. Pati}

\address{Theoretical Sciences Unit and DST unit on Nanoscience
\\Jawaharlal Nehru Centre For Advanced Scientific Research
\\Jakkur Campus, Bangalore 560064, India.}

\date{\today}
\widetext

\begin{abstract}
\parbox{6in}

{We have studied zigzag and armchair graphene nano ribbons (GNRs), described 
by the Hubbard Hamiltonian using quantum many body configuration interaction 
methods. Due to finite termination, we find that the bipartite nature of the 
graphene lattice gets destroyed at the edges making the ground state of the 
zigzag GNRs a high spin state, whereas the ground state of the armchair GNRs 
remains a singlet. Our calculations of charge and spin densities suggest that, 
although the electron density prefers to accumulate on the edges, instead of spin 
polarization, the up and down spins prefer to mix throughout the GNR lattice. 
While the many body charge gap results in insulating behavior for both 
kinds of GNRs, the conduction upon application of electric field is still 
possible through the edge channels because of their high electron density. 
Analysis of optical states suggest differences in quantum efficiency of 
luminescence for zigzag and armchair GNRs, which can be probed by simple
experiments.}
\end{abstract}

\maketitle

\narrowtext

Nanomaterials of carbon of different dimensionalities have been a subject of 
interest over the past few decades due to their potential applications in 
various nanoscale electronic devices\cite{Dresselhaus,Chico,McEuen}. 
Compared to fullerenes and 
carbon nanotubes, which are effectively zero and one dimensional in nature, 
the two dimensional flat monolayer of carbon atoms packed into a honeycomb 
lattice, i.e., graphene, has started gaining prominence very recently owing to 
the recent progress in experimental techniques\cite{Novo1,Novo2,Zhang,Stankovich}. 
Because of its sophisticated low-dimensional electronic properties and huge 
application possibility, it has attracted a big scientific army to explore it 
in various aspects\cite{Meyer,Novo3,Novo4,Katsnelson}. 
   
The size and geometry of the nanoscale carbon systems govern their electronic 
properties. Recent progress in experiments allows to make finite size graphene 
layer, termed as graphene nanoribbons (GNRs) with varying widths, either by 
cutting mechanically exfoliated graphenes\cite{Novo1,Novo2,Zhang} and patterning 
by electron beam lithography\cite{Ozyilmaz}, or by controlling the epitaxial 
growth of graphenes\cite{Berger1,Berger2}. Different possibilities of geometrical 
termination of graphene layer give rise to two different edge geometries, namely, 
zigzag and armchair edges, differing largely in their electronic properties. 
These different edge geometries have been modeled by imposing different boundary 
conditions on Schrodinger's equation within the tight-binding 
limit\cite{Fujita,Nakada,Wakabayashi,Ezawa} or on the Dirac equation for 
two-dimensional massless fermions with an effective speed of 
light\cite{Brey,Sasaki,Abanin,Neto} in previous studies. These also have been 
extensively studied using density functional theory\cite{Cohen,Okada}.
However, the proper many-body description of the GNR is still lacking. To obtain 
new inroads in the rich physics of one atom thick finite graphene material, for 
the first time, we use configuration-interaction (CI) method based on many-electron 
theory with proper inclusion of electron interactions term which plays
crucial role in low-dimensional systems.

Our CI approach with correlation parameters in Hamiltonian obtained semiempirically, 
can directly connect to the real description of the system under investigation.
We study both the zigzag and armchair edges with a fixed width and vary their 
size by translating it in one direction up to a large limit to reduce finite 
size effects. Since, the full many-body CI calculation is not possible for 
very large system, we map the whole system into a complete active space (CAS) 
which captures the low energy states accurately. We have also considered 
the Single CI (SCI) which shows the same qualitative features as obtained from
CAS-CI. 

\begin{figure}
\centering
\includegraphics[scale=0.15, angle=0] {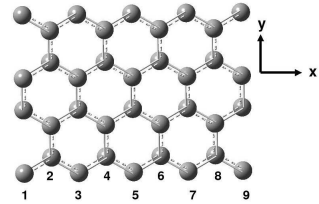}
\caption{The unit cell of GNR. The translations along $x$ and $y$-axis give
the zigzag and armchair edge GNRs. The number of atoms at the zigzag terminal
of an armchair GNR are represented by the integers.}
\end{figure}

We employ Hubbard Hamiltonian for the GNRs,

\begin{eqnarray}
H &=& \sum\limits_{i}\epsilon_{i}a^\dagger{_i}a_{i} \nonumber 
+ \sum\limits_{<i,j>,\sigma}t_{ij}(a^\dagger_{i,\sigma}a_{j,\sigma}+h.c) \nonumber \\
&+& U \sum\limits_{i}n_{i\uparrow}n_{i\downarrow}
\end{eqnarray}
 
\noindent with standard notations. The nearest neighbor hopping integral $t_{ij}$
is considered to be $2.4$ eV which is standard for the C-C bond. We 
consider $U$ as the difference between the first ionization energy and the electron 
affinity for carbon, which comes out to be $\sim 9.66$ eV. Due to finite termination 
of the GNRs, the chemical nature
of the edge atoms changes abruptly because of unsatisfied valence of carbon atoms.
As a result, the lattice deviates from its 
bipartite nature at the edges. So, to differentiate these edge atoms from the bulk, 
we model our system by putting non zero negative onsite energy, $\epsilon_{edge}$
at the edge atoms to mimic hydrogen passivation and $\epsilon_{i}=0$ for 
the bulk. Although, the onsite energies at edges are expected to be modified compared 
to the bulk as a difference between the electronegativities of carbon and 
hydrogen\cite{Ezawa}, they have been neglected in most of the studies 
so far. Moreover, the scanning tunneling microscopy (STM) images of graphite shows 
bright stripes along the edges, suggesting more electron density on the edge atoms 
compared to the bulk\cite{Kobayashi}. This observation motivates us to simulate the
GNRs with negative onsite energies at the edge carbon atoms to create a potential 
well which can trap the electrons on the edges. 
We compare and contrast our model with the bipartite lattice model, where 
all the lattice points are assigned $\epsilon_{i}=0$, both at the tight binding (TB) 
and at the CI level of calculations.

\begin{figure}
\centering
\includegraphics[scale=0.25, angle=270] {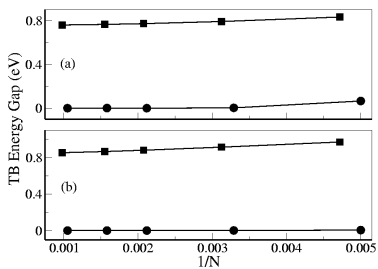}
\caption{The lowest gap in tight binding calculations as a function of inverse 
system size ($1/N$) for both zigzag $(circle)$ and armchair $(square)$ GNRs with 
$\epsilon_{edge}$ equals to $(a)-2.0$ and $(b)0$.}
\end{figure}

Fig.1 displays the GNRs under investigation. The translations along $x$-axis
and $y$-axis produce zigzag and armchair GNRs of different sizes respectively.
We consider the armchair GNRs with $3p$ number of atoms constituting the two zigzag 
terminals, where $p$ $(=3$ in our case$)$ is an integer. In Fig.2, we plot the 
TB $(U=0)$ gap for both types of GNRs with and without 
consideration of edge hydrogen passivation energies, with inverse system size. 
The calculated TB gap reproduces the well established 
previous observations of zigzag being metallic and armchair $(3p)$ being 
semiconducting GNRs\cite{Fujita,Nakada,Wakabayashi,Ezawa,Brey,Sasaki,Abanin,
Cohen,Okada,Reich}. However, inclusion of electron correlations are 
expected to open up a gap reducing the kinetic stabilization.

\begin{table}
\centering
\caption{The ground state energies (E($S_{z}$)) per atom for two different 
system sizes ($N$) for both zigzag and 
armchair GNRs at half filling with different $\epsilon_{edge}(eV)$ values. 
Since all total spin states $(S^{tot})$ have projection into 
$S_{z}^{tot}\le S^{tot}$, the ground state spin is determined by 
the highest $S_{z}^{tot}$ value with ground state energy.}

\begin{tabular*}{2.92in}{|c|c|c|c|c|c|} \hline
\multicolumn{3}{|c|} {Zigzag} &  \multicolumn{3}{|c|} {Armchair} \\ \hline
   & N=304 & N=952 & & N=320 & N=1022 \\ \hline
 $\epsilon_{edge}$ & E($S_{z}$) & E($S_{z}$) & $\epsilon_{edge}$ & E($S_{z}$) & E($S_{z}$) \\ \hline
 -0.2 & -1.1951(4) & -1.1913(4) & -0.2 & -1.2508(0) & -1.2576(0) \\ \hline
 -0.5 & -1.2750(4) & -1.2672(4) & -0.5 & -1.3247(0) & -1.3272(0) \\ \hline
 -1.0 & -1.4046(4) & -1.3914(4) & -1.0 & -1.4492(0) & -1.4446(0) \\ \hline
 -2.0 & -1.6580(4) & -1.6313(4) & -2.0 & -1.6982(0) & -1.6830(0) \\ \hline
\end{tabular*}
\end{table}

Based on the TB wave functions, we perform the CI calculations over the 
electron correlations within many body formalism. The $z$-component of total 
spin $(S_z^{tot})$ is used as a quantum number together with the $N$ number of 
electrons. We vary $S_{z}^{tot}$ from zero to higher values for half-filled 
systems. In CAS-CI, we consider all possible configurations within a small 
energy window (the active space) and obtain the CI matrix of order $4900$.
We have verified our results with $63504$ states in certain cases.
We have also considered singles CI (SCI) calculations by varying the number 
of occupied and unoccupied single particle levels. However, since the CAS-CI 
includes all the configurations within a certain energy cut-off, the resulting 
low-energy states are more size consistent than those from the SCI and unless 
otherwise stated, all the results reported below are obtained using CAS-CI 
approach. For any kind of energy cut off with $S_{z}^{tot}=0$, we find that 
the diagonals of the CI matrix appears in the order of $\frac{NU}{4}$ after 
correction for the levels beyond the cut off.

Our many-body calculations for half-filled GNRs with different choices of 
$\epsilon_{edge}$ value and with varying $S_{z}$ states, lead to results, which 
have been anticipated\cite{Cohen,Hikihara} but not observed till now: that, the 
ground state of the zigzag GNR is always a high spin state, whereas, it remains
in a singlet state for the armchair GNRs. In Table1, the numbers in parenthesis
for various size GNRs are the total ground state spins. 
This observation remains consistent for all nonzero negative 
$\epsilon_{edge}$ values. However, for $\epsilon_{edge}=0$, both the GNRs show 
singlet ground state, which is expected when the lattice is bipartite 
with same number of atoms in the two sublattices\cite{Lieb}. 
These observations are consistent with the gradual 
increase of the system size for a fixed width. To correlate the experimental 
observation\cite{Kobayashi}, we calculate the charge density over all the 
atoms for both zigzag and armchair GNRs with $N=104$, with $\epsilon_{edge}=0$ 
and $-2.0$ (Fig.3). The $\epsilon_{edge}=0$ value shows almost same 
electron densities on all the atoms, whereas non-zero $\epsilon_{edge}$ leads to 
fairly large charge densities on the edge atoms (maxima in Fig.3) for both types of 
GNRs as can be seen from Fig.3. This is the result of unsatisfied coordination of 
the edge atoms, passivated by hydrogen. Interestingly, the charge accumulation at 
the zigzag edges are more than those at the armchair edges. Note that, the minima 
in the charge density plots correspond to the atoms directly connected to the edge 
atoms. This reduction in charge densities on the left and right neighbors of edge 
atoms gives kinetic stability to the extra electrons accumulated at the edge atoms.

\begin{figure}
\centering
\includegraphics[scale=0.65, angle=0] {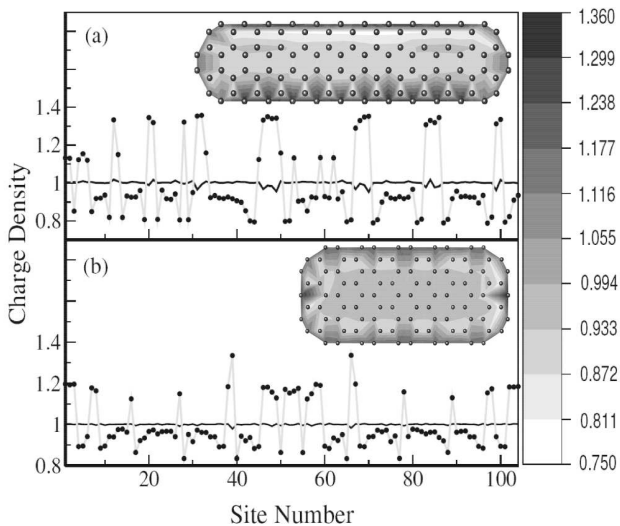}
\caption{Charge density on each and every site for both zigzag $(a)$ and armchair
$(b)$ GNRs with $\epsilon_{edge}=-2.0$ (grey lines with circles) and $0$
(solid lines without any symbol) on the edge atoms of the system with 
$N=104$. Insets present the contour plot of the charge density for the 
corresponding systems with $\epsilon_{edge}=-2.0$.}
\end{figure}

To understand the nature of the spin at the edges, we calculate the spin density
over all the atoms, which suggests that both zigzag and 
armchair GNRs do not prefer any dominant spin at the edges as well as on the
bulk atoms for both zero and non-zero $\epsilon_{edge}$ values. Instead, a proper many 
body consideration shows that, the GNRs prefer to have a mixture of both up and down 
spin density at the edges. This observation clearly contradicts previous 
density functional studies which predict the dominance of up spins and down 
spins on either edges of 
zigzag GNRs\cite{Cohen,Okada}. Those studies conclude this observation as
the property of bipartite lattice with two different sublattice points.
However, the presence of passivating hydrogen atoms on the edges would 
destroy the bipartite nature of the lattice which, however, have been 
completely ignored. Moreover, the one electron 
definition of the exchange correlations in DFT studies cannot capture the 
dominant electron electron interactions in GNRs. The more realistic many body 
calculations infer that, although the electrons have the tendency to accumulate at 
the edges, a net spin polarization of the edges is highly improbable.

Now to investigate the possibility of electron conduction in the different GNRs,
we calculate the charge gap. The many-body charge 
excitation gap is defined as the difference between the energy required to add 
($\mu_+$) and remove ($\mu_-$) electrons from the ground state\cite{Wu},

\begin{eqnarray}
\Delta_{charge}=\mu_{+}-\mu_{-}
\end{eqnarray}

\noindent where $\mu_{+}=E(N+1)-E(N)$ and $\mu_{-}=E(N)-E(N-1)$.
$E(N)$, $E(N+1)$ and $E(N-1)$ are the energies of the half-filled system and
the systems with one extra and one less electron respectively.
The charge gap results from our calculations show Mott insulating behavior
for both zigzag and armchair GNRs even for very large system sizes with 
more than thousand atoms as observed in earlier many body study\cite{Hikihara}. 
This result is expected since the kinetic stabilization
is suppressed by strong electron correlations, which is unexpected within 
one-electron theories. For a half-filled system, with large electron densities 
at the edges, the conduction, however, is still possible through edge channels. 

Motivated by the difference in the spin multiplicities of the ground and 
low-lying excited states between the zigzag and armchair GNRs,
we study their optical properties. We have 
calculated the transverse and longitudinal transition dipole moments, $\mu_T$ 
and $\mu_L$ respectively, for the excitations from the ground state to the 
optically allowed excited states and present $\mu^{2}=\mu_T^2+\mu_L^2$ for 
$N=104$ in Table 2. As can be seen from Table 1, the energy levels in zigzag 
GNRs are more closely spaced than those in the armchair GNRs. Similar to the 
ground state, the first few excited states of zigzag GNRs are also of higher 
spin states. However, due to nonzero $\epsilon_{edge}$, the high spin ground 
states are not degenerate with different $S_z^{tot}$ components. With nonzero 
gap values, these low energy excitations show large absorption cross sections. 
Moreover, the optically allowed emissions from the excited states are expected 
to show very high quantum efficiency (QE) in the luminescence spectroscopy, 
since according to Kasha's rule\cite{Kasha} the spin allowed optical
emission can occur from high spin excited state to the high spin ground state.
In the case of armchair GNRs, however, the singlet ground state is followed by 
a few high spin excited states. This dipole forbidden excited states below 
the allowed optical excitation originate from the strong electron electron 
interactions\cite{Hudson,Sumit}. Thus, it is expected that, the optically excited 
state in armchair GNRs would decay to the low energy dipole forbidden states 
through magnon emission, thereby preventing the radiative transition considerably. 
As a result, the armchair GNRs are expected to show very low QE in luminescence 
spectroscopy. These results are all consistent with the gradual increase of system 
size with fixed width. Thus, from our study on optical properties, we are able to 
propose a simple experimental technique of measuring luminescence to 
differentiate the zigzag and armchair GNRs.

\begin{table}
\centering
\caption{The square of the transition dipole moment $(\mu^2)$ for transitions
from the ground state to the few optically allowed excited states with the
corresponding excitation gaps for the system with $N=104$ in both zigzag and 
armchair GNRs with $\epsilon_{edge}=-2.0$}

\begin{tabular*}{1.94in}{|c|c|c|c|c|} \hline
\multicolumn{2}{|c|} {Zigzag} & & \multicolumn{2}{|c|} {Armchair} \\ \hline
Gap $(eV)$ &   $\mu^{2}$ & & Gap $(eV)$  &    $\mu^{2}$ \\ \hline
0.016    &    2.801    & & 0.906   &   3.770 \\
0.071    &    0.440    & & 0.969   &   0.001 \\
0.073    &    3.069    & & 1.060   &   3.027 \\
0.115    &    0.168    & & 1.136   &   0.002 \\ \hline

\end{tabular*}
\end{table}

While our discussions so far are based on short range Hubbard Hamiltonian,
it is known that the one dimensional conjugated carbon systems are best described 
by long range Coulomb interactions\cite{Sumit}. For graphene class of systems, 
however, even the on-site Hubbard correlations have hardly been 
considered\cite{Fujita,Nakada,Wakabayashi,Ezawa,Brey,Sasaki,Abanin,
Neto,Cohen,Okada,Reich}. There are also suggestions that the Hubbard 
repulsion can at best be $3-4eV$ in graphene\cite{Baskaran}, three times smaller
in magnitude than what is considered for the conventional conjugated 
polyemers\cite{Sumit}. Interestingly, we have verified that all our
observations described earlier are consistent even with $U=3$eV. However,  
to test whether longer range interactions exist in such systems, we have 
investigated these GNRs with long-range Coulombic interactions using PPP Hamiltonian 
within two decay
profile schemes originally given by Ohno and by Mataga-Nishimoto\cite{Ohno}. 
Our results indicate that 
for both these potential profiles, the ground state has large contribution from 
either high-energy singles configuration (in case of SCI) or configurations with 
two or three electron excitations within the energy window for the CAS-CI calculations. 
Contributions from these configurations in fact raise the energy of the state
with finite magnetic moment. This hence results in both zigzag and armchair GNRs 
having a singlet ground state, while the lowest excitations become magnetic.

In summary, the quantum many-body configuration interaction method captures 
the low-energy properties of the nano scale systems like graphene and our 
development of CI method allows us to handle fairly large systems which can 
effectively be considered as infinite lattice. Within the Hubbard model, the
ground state of the zigzag GNRs is a nonzero spin state, while for armchair 
GNRs it remains a singlet. Though the hydrogen passivation of the edge atoms
leads to higher charge density at the edges, the electronic correlations mix
the up spin and down spin throughout the GNR lattices, instead of making the 
edges spin polarized. We propose that, the zigzag GNRs can be differentiated 
from armchair GNRs from the higher quantum efficiency of luminescence
originating from their magnetic ground and low lying excited states.
It would also give an indication of the importance of the long range Coulomb 
interactions in these class of systems. Our findings on GNRs suggest 
rich low-energy physics and provoke further studies within many body limit 
for these class of low-dimensional systems. 

Acknowledgement: SD acknowledges the research support from CSIR and SKP acknowledges 
the research grant from DST and CSIR, Govt. of India.

\end{document}